# The Nature of Bonding in Bulk Tellurium Composed of One-Dimensional Helical Chains

Seho Yi[†], Zhili Zhu[§], Xiaolin Cai[§], Yu Jia[*,‡,§], Jun-Hyung Cho[*,†]

[†] Department of Physics, Hanyang University, 222 Wangsimni-ro, Seongdong-Ku, Seoul 04763, Korea

[‡] International Laboratory for Quantum Functional Materials of Henan, and School of Physics and Engineering, Zhengzhou University, Zhengzhou 450001, China

[§] Key Laboratory for Special Functional Materials of Ministry of Education, and School of Physics and Electronics, Henan University, Kaifeng 475004, China

*Supporting Information*

**ABSTRACT:** Bulk tellurium (Te) is composed of one-dimensional (1D) helical chains which have been considered to be coupled by van der Waals (vdW) interactions. However, based on first-principles density-functional theory calculations, we here propose a different bonding nature between neighboring chains: i.e., the helical chains made of normal covalent bonds are connected together by coordinate covalent bonds. It is revealed that the lone pairs of electrons of Te atom participate in forming coordinate covalent bonds between neighboring chains. Therefore, each Te atom behaves as both electron donor to neighboring chains and electron acceptor from neighboring chains. This ligand-metal-like bonding nature in bulk Te results in the same order of bulk moduli along the directions parallel and perpendicular to the chains, contrasting with the large anisotropy of bulk moduli in vdW crystals. We further find that the electron effective masses parallel and perpendicular to the chains are almost the same each other, consistent with the observed nearly isotropic electrical resistivity. It is thus demonstrated that the normal/coordinate covalent bonds parallel/perpendicular to the chains in bulk Te lead to a minor anisotropy in structural and transport properties, distinct from a strong anisotropy observed in the typical two-dimensional (2D) vdW materials such as graphite and $MoS_2$.

Due to its multivalency character, tellurium (Te) exhibits a wide variety of stable structures under pressure, which contain many coordination numbers ranging from 2 to 8[1]. The most stable structure of Te has a trigonal crystal lattice at ambient pressure[1a], which consists of 1D helical chains with a coordination number of two. For this bulk-Te structure, the general consensus is that the nearest neighboring atoms along the chains are linked through strong covalent bonds, while the next nearest neighboring atoms between the chains are coupled by weak van der Waals (vdW) interactions (see Figure 1). This anisotropic bonding nature in bulk Te is naturally expected to exhibit drastically different electrical transport properties along the directions parallel and perpendicular to the chains. However, surprisingly, earlier experimental studies[2] reported that the electrical resistivity of bulk Te measured at room temperature was $\rho_\parallel = 0.26$ Ω·cm and $\rho_\perp = 0.51$ Ω·cm parallel and perpendicular to the chains, respectively. Recently, these somewhat isotropic transport behaviors in bulk Te were also observed in large area, high-quality 2D tellurium, where the ratio $\rho_\perp/\rho_\parallel$ is only ~1.13[3]. It is, however, noted that the typical 2D layered materials with vdW interlayer interactions, such as graphite and $MoS_2$, exhibit a huge anisotropic behavior with $\rho_\perp/\rho_\parallel \approx 0.001$[4], where $\rho_\perp$ ($\rho_\parallel$) represents the electrical resistivity across (along) the layers. This result indicates that the transport along the weak vdW bound direction is significantly slower than that along the chemical bond direction. Thus, the observed nearly isotropic transport properties of bulk Te in the directions parallel and perpendicular to the chains are unlikely to represent a 1D vdW crystal generally accepted so far[2,3,5], but invoke other binding mechanism between the helical chains. In this communication, using the systematic density-functional theory (DFT) calculations with the local, semilocal, and meta-semilocal exchange-correlation functionals, we demonstrate that the interaction between neighboring helical chains is characterized by coordinate covalent binding with lone-pair electrons, therefore enabling each Te atom to attain four coordinate covalent bonds between neighboring chains (see Figure 1). As a result, the electron effective masses parallel and perpendicular to the chains are found to be very similar as ~0.11 $m_0$ and ~0.13 $m_0$, respectively. Thus, our findings not only propose a new bonding nature in bulk Te with a coordinate covalent bonding between the helical chains but also provide an explanation for the observed isotropic transport behaviors.

Our DFT calculations were performed using the Vienna ab initio simulation package (VASP) code with the projector augmented wave method[6]. For the treatment of exchange-correlation energy, we employed various functionals including the local density approximation (LDA) functional of Ceperley-Alder (CA)[7], the generalized-gradient approximation (GGA) functional of Perdew-Burke-Ernzerhof (PBE)[8], and the meta-GGA functional of SCAN[9]. A plane wave basis was employed with a kinetic energy cutoff ($E_{cut} = \hbar^2 G_{MAX}^2/2m$) of 500 eV. To calculate the charge density, we accommodated its Fourier components within a cutoff of $2G_{MAX}$. The k-space integration was done with the 25×25×19 meshes in the Brillouin zone. All atoms were allowed to relax along the calculated forces until all the residual force components were less than 0.001 eV/Å.

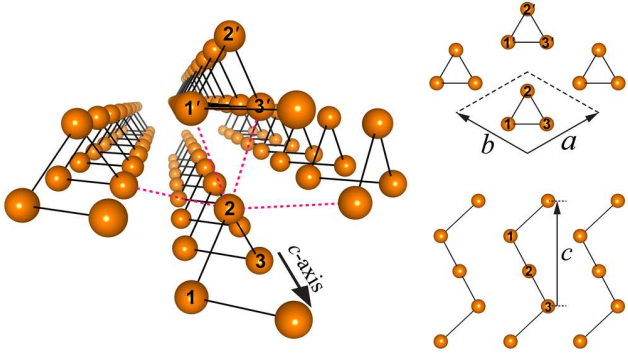

Figure 1. Perspective, top, and side views of the equilibrium structure of bulk Te, obtained using the meta-GGA functional of SCAN. The bonds between nearest neighbors (next nearest neighbors) are represented by the solid (dashed) lines in the perspective view. The lattice parameters are represented by $a$, $b$, and $c$.

We begin to calculate the total energy $E_{Te}$ of bulk Te as a function of volume $V$ using various exchange-correlation functionals. Here, the minimum energy for each volume is determined by optimizing the axial ratio $a/c$. By fitting this $E_{Te}$–$V$ curve to the Birch-Murnaghan equation of state[10], we obtain the equilibrium lattice constants $a_0$ and $c_0$ and bulk modulus $B$ (see the Section 1 in the Supporting Information). The results are summarized in Table 1 together with the bond lengths $d_{NN}$ and $d_{NNN}$. We find that LDA-CA (GGA-PBE) underestimates (overestimates) the equilibrium lattice constant $a_0$ compared to the experimental value[11] by 3.8 (1.3) %, while the meta-GGA-SCAN value is in good agreement with experiment. Meanwhile, the equilibrium lattice constant $c_0$ is reasonably well predicted by all the exchange-correlation functionals. For bulk modulus, meta-GGA-SCAN also agrees well with experiment (see Table 1). Figure 1 shows the equilibrium structure of bulk Te obtained using meta-GGA-SCAN. There are two nearest neighbors (NNs) along the chain and four next nearest neighbors (NNNs) between the chains, giving rise to a coordination number of 6. The calculated bond length $d_{NN}$ is 2.91, 2.89, and 2.87 Å for LDA-CA, GGA-PBE, and meta-GGA-SCAN, respectively, while $d_{NNN}$ 3.31, 3.50, and 3.45 Å. Since the magnitude of $d_{NNN}$ is close to the interlayer distance in the typical 2D vdW materials (e.g., 3.34 Å in graphite[12] and 3.49 Å in MoS$_2$[13]), it has been presumed that the interchain interaction in bulk Te would be of vdW type[2,3,5]. However, it is noticeable that the sum of the vdW radii of two Te atoms amounts to ~4.12 Å[14], sufficiently larger than our calculated values of $d_{NNN}$. As discussed below, the relatively longer bond length of $d_{NNN}$ compared to $d_{NN}$ is attributed to the relatively weaker coordinate covalent bonding character between neighboring chains than the normal covalent one along the chains.

**Table 1. Calculated lattice constants and bulk modulus of bulk Te, in comparison with the experimental values.**

|  | $a$ | $c$ | $B$ | $d_{NN}$ | $d_{NNN}$ |
|---|---|---|---|---|---|
| LDA-CA | 4.28 | 5.93 | 38 | 2.91 | 3.31 |
| GGA-PBE | 4.51 | 5.96 | 18 | 2.89 | 3.50 |
| Meta-GGA-SCAN | 4.45 | 5.93 | 18 | 2.87 | 3.45 |
| Experiment[11] | 4.45 | 5.93 | 19 | – | – |

The calculated bond length $d_{NN}$ ($d_{NNN}$) between nearest neighbors (next nearest neighbors) is also given. The unit is in Å.

Figure 2a shows the total charge density $\rho_{Te}$ of bulk Te, obtained using meta-GGA-SCAN. Obviously, it is seen that $\rho_{Te}$ represents not only the covalent character for the NN bonds along the helical chains but also the relatively weaker covalent character for the NNN bonds between the chains. Note that each bond has a saddle point of charge density at its midpoint [see Figure 2a], similar to the C-C covalent bond in diamond[15]. Here, the charge density at

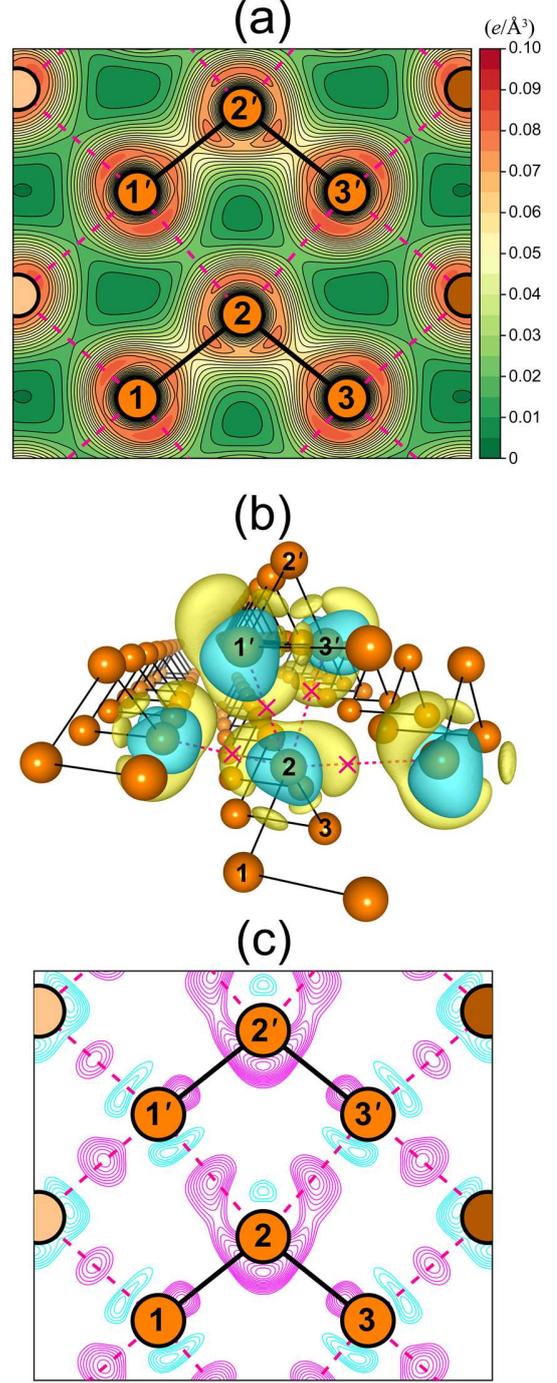

Figure 2. (a) Total charge density $\rho_{Te}$ of bulk Te. The charge density differences $\Delta\rho$ and $\Delta\rho'$, defined in the text, are also given in (b) and (c), respectively. In (a), the first line is drawn at $5\times10^{-5}$ $e$/Å$^3$ and the contour spacing is $5\times10^{-3}$ $e$/Å$^3$. In (b), $\Delta\rho$ is drawn with an isosurface of $\pm5\times10^{-3}$ $e$/Å$^3$. Here, the accumulated (depleted) electrons are represented by the yellow (blue) isosurface, and lone pairs of the Te$_2$ atom and its NNNs (Te$_{1'}$ and Te$_{3'}$) are marked (×). In (c), the accumulated (depleted) electrons are represented by the magenta (cyan) lines, where the first line is drawn at $1.5\times10^{-3}$ ($-1.5\times10^{-3}$) $e$/Å$^3$ and the contour spacing is $2.5\times10^{-4}$ ($-2.5\times10^{-4}$) $e$/Å$^3$.

**Table 2. Calculated elastic constants $C_{ij}$, shear modulus $G$, Young's modulus $E$, and Poisson's ratio $v$ using meta-GGA-SCAN, in comparison with previous GGA-PBE calculation and experiment.**

|  | $C_{11}$ | $C_{12}$ | $C_{13}$ | $C_{14}$ | $C_{33}$ | $C_{44}$ | $G$ | $E$ | $v$ |
|---|---|---|---|---|---|---|---|---|---|
| Meta-GGA-SCAN | 28 | 6 | 19 | 9 | 69 | 30 | 16 | 38 | 0.19 |
| GGA-PBE[19] | 29 | 7 | 20 | – | 63 | 29 | – | – | – |
| Experiment[20] | 33 | 8 | 26 | 12 | 72 | 31 | – | – | – |

The unit of elastic constants, shear modulus, and Young's modulus is GPa.

the midpoint of the NN and NNN bonds is 0.06 and 0.02 $e$/Å$^3$, respectively. To explore the more-detailed bonding character of bulk Te, we calculate the charge density difference defined as $\Delta\rho = \rho_{Te} - \rho_{atoms}$, where $\rho_{atoms}$ is the superposition of the atomic charge densities. As shown in Figure 2b, $\Delta\rho$ shows not only a charge accumulation in the middle of the NN bond but also the presence of two lone pairs per each Te atom, while accompanying a depletion of charge in some regions around Te atoms. It is thus likely that an NN bond along the chains is characterized as the normal covalent bond, but an NNN bond between the chains as the coordinate covalent bond where each lone pair of electrons participates in forming a bond with the Te atom in a neighboring chain. These two different bonding natures between NNs and NNNs are well represented by the calculated bond lengths: i.e., $d_{NNN}$ is longer than $d_{NN}$ by 0.4~0.6 Å (see Table 1) due to its weak coordinate covalent bonding. We note that the nonmetallic character among group-VI elements is weakened in the order of O > S > Se > Te, leading to a complete metallic character of Po. Here, Te tends to have the dual characteristics of both non-metal and metal with ligand-metal-like bonding. Interestingly, such unique bonding features of Te can be seen in Figure 2b: i.e., the Te$_2$ atom with two lone pairs behaves as electron donor to NNNs as well as electron acceptor from NNNs (Te$_{1'}$ and Te$_{3'}$).

To estimate the strength of the interaction between the helical chains in bulk Te, we calculate the interchain binding energy, defined by $E_b = E_{chain} - E_{Te}$, where $E_{chain}$ is the total energy of an isolated helical chain. We find that LDA-CA (GGA-PBE) gives $E_b$ = 0.490 (0.173) eV/atom, which is larger (smaller) than that (0.255 eV/atom) obtained using meta-GGA-SCAN. This overestimation (underestimation) of $E_b$ in LDA-CA (GGA-PBE) is well represented among the calculated values of $d_{NNN}$, which are in the order of GGA-PBE (3.50 Å) > meta-GGA-SCAN (3.45 Å) > LDA-CA (3.31 Å). It is noted that our calculated meta-GGA-SCAN interchain binding energy of bulk Te is much larger than the observed interlayer binding energy of graphite[16] ranging between 0.031 and 0.052 eV/atom. Moreover, the bonding natures between bulk Te and such common vdW materials should be distinguished from each other. As shown in Figure 2c, the covalent bonding character between neighboring Te chains can be seen from the charge density difference $\Delta\rho' = \rho_{Te} - \rho_{chain}$, where $\rho_{chain}$ is the superposition of the charge densities of isolated chains. Here, $\Delta\rho'$ clearly shows charge accumulation in the middle regions between the chains, not supporting vdW interactions between the helical chains.

Next, we study the elastic properties of bulk Te which can reflect the proposed covalent bonding nature. Figure 3 shows the variation of $E_{Te}$ as a function of the lattice parameter ratios, $a/a_0$ and $c/c_0$, obtained using meta-GGA-SCAN that predicts well the experimentally measured lattice constants (see Table 1). By using the Birch-Murnaghan equation of state, we fit the $E_{Te}$ vs. $a$ ($c$) curve to obtain a bulk modulus of 18 (65) Pa. The somewhat larger bulk modulus along the $c$ axis compared to the $a$ axis indicates that the normal covalent bonding parallel to the chains gives relatively stiffer with changing the lattice parameter $c$, compared to the case of the coordinate covalent bonding perpendicular to the chains. However, since the two bulk moduli $B_a$ and $B_c$ perpendicular and parallel to the chains are in the same order of magnitude, we can say that the elastic anisotropy of bulk Te is minor, contrasting with the typical 2D vdW materials such as graphite, where the bulk moduli along the in-plane and out-of-plane directions were observed to be anisotropic with a large ratio of ~35[17]. Thus, the small ratio of $B_c/B_a \approx 3.6$ in bulk Te also does not support a vdW bonding picture between the helical chains. To estimate the stiffness of bulk Te against the external strain, we further calculate the elastic constants which are defined as the second-order derivatives of $E_{Te}$ with respect to the infinitesimal strain tensor[18]. For bulk Te, there are six independent elastic constants $C_{11}$, $C_{12}$, $C_{13}$, $C_{14}$, $C_{33}$, and $C_{44}$. Using the Voigt-Reuss-Hill approximation (see the Section 2 in the Supporting Information), we obtain other elastic parameters such as shear modulus $G$, Young's modulus $E$, and Poisson's ratio $v$. Our results for $C_{ij}$, $G$, $E$, and $v$ are given in Table 2. It is seen that the present values of $C_{ij}$ are in good agreement with previous theoretical[19] and experimental[20] data.

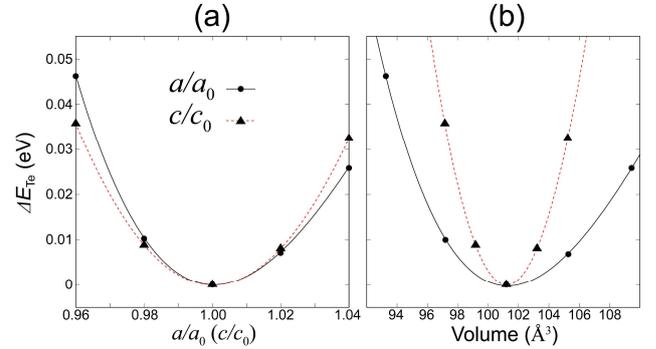

Figure 3. (a) Total energy difference (in eV per Te atom) as a function of the lattice parameter ratios, $a/a_0$ and $c/c_0$, obtained using meta-GGA-SCAN. In (b), $\Delta E_{Te}$ vs $a/a_0$ ($c/c_0$) is converted to $\Delta E_{Te}$ vs volume $V$. Here, $V$ is equal to $\sqrt{3}a^2c_0/2$ ($\sqrt{3}a_0^2c/2$) in the direction perpendicular (parallel) to the chains.

Since the proposed normal/coordinate covalent bonding natures parallel/perpendicular to the chains have similar charge characteristics (see Figure 2a), it is natural to expect isotropic electrical transport properties as observed in experiments[2,3]. Figure 4 shows the band structure of bulk Te, obtained using meta-GGA-SCAN. We find a semiconducting feature with a band gap of $E_g$ = 0.08 eV. This theoretical band gap is much underestimated compared to the experimental value[21] of 0.33 eV. In order to properly predict the measured band gap, we perform the hybrid DFT calculation with the HSE functional[22], where the gap size depends on the magnitude of α controlling the amount of exact Fock exchange energy[23]. We find that, when the HSE functional with α = 0.125 is used, the band gap increases to $E_g$ = 0.32 eV (see Figure 4), close to the experimental data[21]. Meanwhile, the standard HSE calculation with α = 0.25 is found to give $E_g$ = 0.54 eV. On the basis of the meta-GGA-SCAN (HSE with α = 0.125) band structure of bulk Te, we estimate the electron effective masses parallel and perpendicular to the chains as 0.05 (0.11) $m_0$ and 0.11 (0.13) $m_0$, respectively. These results indicate very isotropic electrical transport properties of bulk Te, consistent with the measurements[2,3] of nearly isotropic resistivity parallel and perpendicular to the chains.

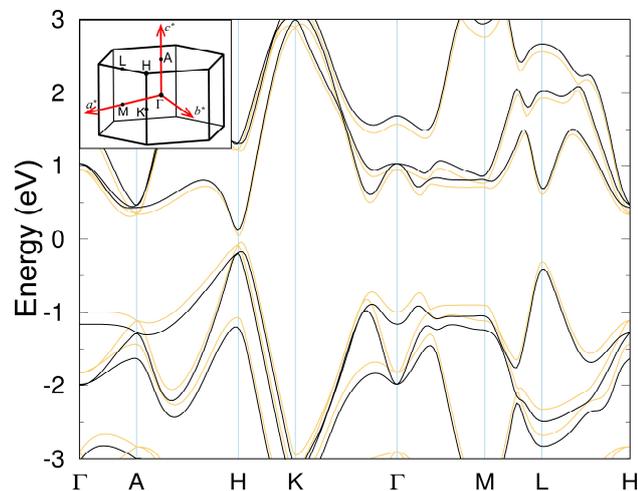

Figure 4. Calculated band structures of bulk Te using the meta-GGA-SCAN functional (bright lines) and the HSE functional with α = 0.125 (dark lines). The Brillouin zone is drawn in the inset.

In summary, our first-principles DFT study of bulk Te has demonstrated that the helical chains made of normal covalent bonds are bound with each other through coordinate covalent bonds. We revealed that the coordinate covalent bonds between neighboring chains are formed by lone pairs of electrons of Te atoms. These covalent bonding characters along the directions parallel and perpendicular to the chains were found to give rise to a minor anisotropy in the structural and transport properties of bulk Te, consistent with experiments[2,3]. Our findings not only elucidate that the nature of binding between neighboring helical chains in bulk Te is characterized as a coordinate covalent bonding rather than the so-far accepted vdW interactions[2,3,5], but also have important implications for understanding the physical properties of Te layers which have attracted much attention recently as a new 2D material for electronic and optical devices[3,5b,5c,24,25].

## ASSOCIATED CONTENT

### Supporting Information

The Supporting Information is available free of charge on the ACS Publications website, including the detailed calculation methods for the elastic properties of bulk Te.

## AUTHOR INFORMATION

### Corresponding Author

*chojh@hanyang.ac.kr
*jiayu@zzu.edu.cn

### Author Contributions

Y.J. and J.H.C designed and supervised the work. S.Y., X.C., Z.Z., and H.J.K performed the theoretical calculations. Y.J, and J.H.C contributed to data interpretation and presentation. Z.Z. and J.H.C. wrote the manuscript. All authors contributed to the scientific discussion and manuscript revisions.

### Notes

The authors declare no competing financial interests.

## ACKNOWLEDGMENT

This work was supported by the National Research Foundation of Korea (NRF) grant funded by the Korea Government (Grants No. 2015R1A2A2A01003248 and No. 2015M3D1A1070639). Y.J is supported by the National Basic Research Program of China (Grant No. 11774078). Calculations were performed by the KISTI supercomputing center through the strategic support program (KSC-2017-C3-0041) for supercomputing application research.

# Supporting information for "The Nature of Bonding in Bulk Tellurium Composed of One-Dimensional Helical Chains"

Seho Yi[†], Zhili Zhu[§], Xiaolin Cai[§], Yu Jia[*,‡,§], Jun-Hyung Cho[*,†]

[†] Department of Physics, Hanyang University, 222 Wangsimni-ro, Seongdong-Ku, Seoul 04763, Korea

[‡] International Laboratory for Quantum Functional Materials of Henan, and School of Physics and Engineering, Zhengzhou University, Zhengzhou 450001, China

[§] Key Laboratory for Special Functional Materials of Ministry of Education, and School of Physics and Electronics, Henan University, Kaifeng 475004, China

### 1. Equilibrium lattice parameters and bulk modulus of bulk tellurium

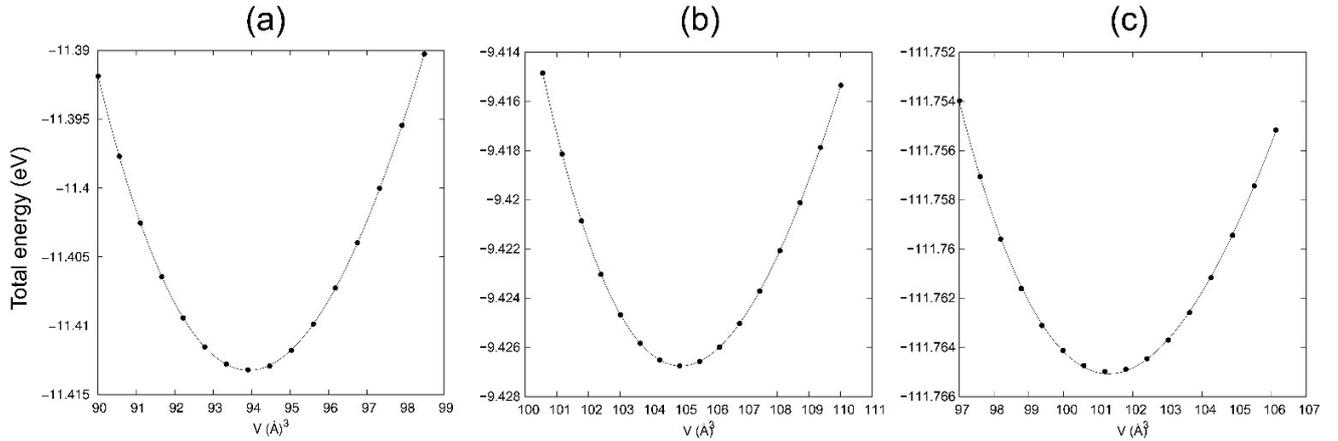

Fig. S1 Calculated equilibrium lattice parameters and bulk modulus using (a) LDA-CA, (b) GGA-PBE, and (c) Meta-GGA-SCAN functionals.

The calculated total energy $E$ as a function of the unit-cell volume $V$ were fitted to the third-order Birch-Murnaghan equation of state (Eq. 1) to obtain the equilibrium volume $V_0$ and bulk modulus $B$ at 0 K and 0 GPa[1].

$$E(V) = E_0 + \frac{9V_0 B}{16}\left\{\left[\left(\frac{V_0}{V}\right)^{\frac{2}{3}} - 1\right]^3 B' + \left[\left(\frac{V_0}{V}\right)^{\frac{2}{3}} - 1\right]^2 \left[6 - 4\left(\frac{V_0}{V}\right)^{\frac{2}{3}}\right]\right\}$$ (Eq. 1)

Here, $B'$ is the derivative of bulk modulus with respect to pressure. At each volume, the $a/c$ ratio and all internal atomic position were fully relaxed.

### 2. Elastic constants and various structural parameters of bulk tellurium

According to the Hooke's law, the elastic constants can be obtained from the deformation of a solid caused by the stress. The elastic constants $C_{ijkl}$ with respect to the finite strain variables are defined as

$$C_{ijkl} = \left(\frac{\partial \sigma_{ij}(X)}{\partial \varepsilon_{kl}}\right)\bigg|_x$$ (Eq. 2)

where $\sigma_{ij}$ is the stress applied to a solid, $\varepsilon_{kl}$ the strain, and $X$ ($x$) the coordinates before (after) deformation. In Eq. 2, we used the Voigt notation where $xx$, $yy$, $zz$, $yz$, $zx$, and $xy$ are replaced by 1, 2, 3, 4, 5, and 6, respectively. The bulk Te has the trigonal-trapezoidal point symmetry with the 32 point symmetry in Hermann-Mauguin notation. Crystals in the rhombohedral class (including trigonal-trapezoidal point symmetry) have six independent elastic constants $C_{11}$, $C_{12}$, $C_{13}$, $C_{14}$, $C_{33}$, and $C_{44}$[2,3],

$$C = \begin{pmatrix} C_{11} & C_{12} & C_{13} & C_{14} & 0 & 0 \\ C_{12} & C_{11} & C_{13} & -C_{14} & 0 & 0 \\ C_{13} & C_{13} & C_{33} & 0 & 0 & 0 \\ C_{14} & -C_{14} & 0 & C_{44} & 0 & 0 \\ 0 & 0 & 0 & 0 & C_{44} & C_{14} \\ 0 & 0 & 0 & 0 & C_{14} & C_{66} \end{pmatrix} \quad \text{(Eq. 3)}$$

where $C_{66} = (C_{11} - C_{12})/2$. The necessary and sufficient elastic-stability conditions of the crystals in the rhombohedral class are[3]

$$C_{11} > |C_{12}|,\ C_{44} > 0,\ C_{13}^2 < \tfrac{1}{2} C_{33}(C_{11} + C_{12})$$
$$C_{14}^2 < \tfrac{1}{2} C_{44}(C_{11} - C_{12}) \equiv C_{44} C_{66} \quad \text{(Eq. 4)}$$

In our DFT calculation with the meta-GGA-SCAN functional, these conditions are satisfied. Furthermore, the other elastic parameters such as bulk modulus ($B$) and shear modulus ($G$) can be calculated. In the rhombohedral class, $B$ and $G$ are given by Voigt[4] and Reuss[5] as follows:

$$B_V = \tfrac{1}{9}[2C_{11} + 2C_{12} + 2C_{13} + C_{33}],\ G_V = \tfrac{1}{15}[2C_{11} - C_{12} - 2C_{13} + C_{33} + 6C_{44} + 3C_{66}]$$
$$B_R = [2S_{11} + 2S_{12} + 2S_{13} + S_{33}]^{-1},\ G_R = \tfrac{1}{15}[8S_{11} - S_{12} - 2S_{13} + 4S_{33} + 6S_{44} + 3S_{66}]^{-1} \quad \text{(Eq. 5)}$$

where $S_{ij}$ are elastic compliances constants. The values of elastic compliances constants can be obtained through an inversion of the elastic constant matrix, $S = C^{-1}$. Hill[6] recommended that a practical estimate of the $B$ and $G$ were the arithmetic averages of the two bound values: $B = (B_V + B_R)/2$ and $G = (G_V + G_R)/2$. The Young's modulus ($E$) and the Poisson's ratio ($v$) can be calculated as $E = 9BG/(3B + G)$ and $v = (3B - 2G)/2(3B + G)$, respectively.